  \documentstyle[preprint,eqsecnum,aps]{revtex}
\def\btt#1{{\tt$\backslash$#1}}
\def\BibTeX{\rm B{\sc ib}\TeX}
\begin{document}
\draft
\preprint{HEP/123-qed}
\title{Charge Fluctuations on Membrane Surfaces in Water}
\author{Rebecca Menes, Philip Pincus\cite{otheraddress} and Bean Stein}
\address{Materials Research Laboratory, University of California 
Santa Barbara, CA 93106}
\date{\today}
\maketitle
\begin{abstract}
We generalize the predictions for attractions between over-all neutral surfaces induced
by charge fluctuations/correlations to non-uniform systems
that include {\it dielectric discontinuities}, as is the case
for mixed charged lipid membranes in an aqueous solution. We show that the
induced interactions depend in a non-trivial way on the dielectric constants
of membrane and water and show different scaling with distance depending on
these properties. The generality of the calculations also allows us to predict 
under which dielectric conditions the interaction will change sign and
become repulsive.
\end{abstract}

\narrowtext

\section{Introduction}
\label{sec:Introduction}

In recent years there has been a growing interest in electrostatic
systems that are dominated by ion fluctuations and ion
distributions around larger charged objects. In some of these systems one
finds attraction between like charged objects\cite{poly} and direct
{\it electrostatic} contributions in systems that are over-all
{\it neutral}\cite{sam,andy}.  

In this paper we will
generalize some theoretical results for systems of neutral surfaces
(membranes) that nontheless interact electrostatically via ion
fluctuations and correlations. These predictions are relevant to
the experimental work done both on biological systems and on artificial
systems where charges are introduced in order to improve membrane
characteristics. Examples are the charged membranes in membrane-DNA
complexes\cite{safinya} used for gene transvection and the
formation of equilibrium bilayer vesicles from mixed charged
lipids\cite{kaler}. 

Recently it has been shown\cite{sam} that charge fluctuations can
lead to attractions between over-all neutral surfaces. 
However, the system treated was the somewhat
artificial case of uniform layers where the interacting surfaces separate
regions of the {\it same} dielectric. In this paper we specifically focus on
the role of the dielectric discontinuities in systems of lipid membranes in
an aqueous solution and how they affect these interactions.  In Sec.\
\ref{sec: 2-d} we introduce a model system for the membrane which includes
two surfaces charged with both positive and negative mobile ions
(charged lipid heads at the bilayer surface) that are over-all
neutral. The system is treated within the Debye-H\"uckel
model\cite{landau,sambook} for a two-dimensional salt
solution\cite{sam,vela}. We calculate the interaction between these two
surfaces resulting from the fluctuations and correlations of the mobile
charges, and find that the resulting attraction
 depends in a non-trivial way on the {\it
dielectric discontinuity} between lipid and water.  

\section {Interaction Between Two Salty Surfaces}
\label{sec: 2-d}

In this section we calculate the effective interaction between two surfaces
that contain mobile charges but are over-all neutral. This is a model
system for mixed charged lipid membranes\cite{kaler} or for membranes that
are very highly charged to the extent that their counter-ions are restricted
to a near-by layer that is thin enough to be considered as a two-dimensional
surface. Pincus and Safran \cite{sam} have calculated this interaction
within the Debye-H\"uckel approximation for a uniform system, {\it i.e.} a
system with no dielectric discontinuities. We will follow their method, 
while introducing the dielectric contributions to this model. 

\subsection {Model}
\label {subsec: 2-d model}

The Debye-H\"uckel model is an expansion of the 
energy to second order in the charge density fluctuations\cite{tombook} and
includes both the electrostatic and entropic contributions due to these
fluctuations:
\begin{equation}
\delta H\!=\! \int\! d\rho d\rho\;'\left
[{1\over 2}\sum_{i=1,2}\left({\delta(\vec\rho\!-\!\vec\rho\; ')\over
\sigma_0}+\phi(\vec\rho\!-\!\vec\rho
\;',z\!=\!0)\!\right)\!\delta\sigma_i(\vec\rho)\delta\sigma_i(\vec\rho
\;')+\phi(\vec\rho\!-\!\vec\rho
\;',z\!=\!d)\delta\sigma_1(\vec\rho)\delta\sigma_2(\vec\rho\; ')\right ]
\label{hamiltonian}
\end{equation}

The self energy of each of the surfaces separately is given by the first two
terms while the third term is the interaction term between charges on the
different surfaces. $\sigma_{1,2}$ are the charge densities on
the surfaces (the index $i=1,2$ denotes the surface number) while
$\rho$ is the in-plane coordinate and
$z$ is the coordinate perpendicular to the surface. The first term ($\delta$
function) is the entropic contribution from the charge density fluctuations
in both surfaces.  In this expression we have assumed, for the sake of
simplicity and without taking away from the generality of the treatment,
that the charge fluctuations are only due to density fluctuations of {\it
one} type of charge while the other sign does not fluctuate and
therefore does not contribute to the free energy to this order. Thus the
entropic contribution can be written in terms of the total
charge density fluctuations on each surface , where
$\sigma_0$ is the average charge density of each species (separately). The
electrostatic contributions,
$\phi$, both between charges in the same surface ($z=0$) and between
charges on the opposite surfaces ($z=d$) are not trivial because of the
dielectric discontinuities that are formed by these surfaces (fig 1). The
discontinuities reflect the fields thus creating image charges in the
region outside the membrane\cite{jackson}. Because this system has two such
discontinuities on either side of the membrane, each image charge is
reflected over and over again so that we have an infinite number of charges
over which to sum when calculating the potential. We require expressions both for the
interaction between charge fluctuations in the same surface (they will also
contribute to the inter-surface interaction via the reflections) and
fluctuations on opposite surfaces.  The interaction potential of two charges
that are in the same surface is:
\begin{equation}
\phi(\vec\rho-\vec\rho
\;',z=0)={e^2\over\bar\epsilon}\left({1\over|\vec\rho-\vec\rho
\;'|}+{\epsilon_m\over
\bar\epsilon}
\sum_{n=1}^{\infty}u^{2n-1}{1\over
\sqrt{|\vec\rho-\vec\rho\; '|^2+(2nd)^2}}\right)
\label{z=0 interaction}
\end{equation}

While the interaction energy for two charges on the two different sides of
the membrane is given by:
\begin{equation}
\phi(\vec\rho-\vec\rho
',z=d)={e^2\epsilon_m\over\bar\epsilon^2}
\sum_{n=1}^{\infty}
u^{2n-2}{1\over
\sqrt{|\vec\rho-\vec\rho\; '|^2+((2n-1)d)^2}}
\label{z=d interaction}
\end{equation}

Here $\epsilon_{w,m}$ are the dielectric constants of water and membrane
lipid respectively,
$\bar\epsilon={\epsilon_w+\epsilon_m\over 2}$,
$u={\epsilon_m-\epsilon_w\over\epsilon_m+\epsilon_w}$ and
$d$ is the membrane thickness. 

The sums in Eq.\ref{z=0 interaction} and \ref{z=d interaction} are easily performed if we
use the identity $\int\exp^{-q z}J_0(qr)dq={1\over\sqrt{r^2+z^2}}$ to transform
them into simple geometric series. The resulting energy in momentum space 
has the form:

\begin{equation}
\delta H=\sum_q\left[{1\over
2}\left(|\delta\sigma_1(q)|^2+ |\delta\sigma_2(q)|^2\right)A(q)+
\delta\sigma_1(q)\delta\sigma_2(q)B(q)\right]
\label{q space hamiltonian}
\end{equation}

The  coefficients are: $A(q)=\left({1\over\sigma_0}+{2\pi <l>\over
q}+{2\pi \delta l\over q}{\exp(-2q d)\over 1-u^2\exp(-2qd)}\right)$ and $B
(q)={2\pi l_m\over q}{\exp(-q d)\over1-u^2\exp(-2 q d)}$. Here we have
defined three different ``Bjerrum lengths":
$<\!\!l\!\!>={e^2\over\bar\epsilon k_BT}$,
$\delta l=<\!\!l\!\!>2{ (\epsilon_m-\epsilon_w)\epsilon_m\over \bar\epsilon^2}$
and
$l_m=<\!\!l\!\!>{\epsilon_m\over\bar\epsilon}$.

At this point it is worth noting the differences between this expression
and that which is found for the uniform case\cite{sam} of no dielectric
variations: The differences are expressed through the various effective
Bjerrum lengths. In the uniform case there is only one such length scale
which would be equal to $<\!\!l\!\!>$ where $\bar\epsilon=\epsilon$. In that case
$l_m=<\!\!l\!\!>=l_B$ and
$\delta l=0$. Hence the differences enter not only in the way they change
the interaction amplitude through $l_m$ and $<\!\!l\!\!>$, but also by 
adding an
additional interaction term that is $d$ dependent, but which is also proportional to the dielectric difference,
$\epsilon_m-\epsilon_w$, through $\delta l$, and affects the resulting interaction 
in a non-trivial way, as will be seen below.\cite{text1} 

The Gibbs free energy for these fluctuations is now given by the logarithm
of the partition function:

\begin{equation}
{G\over k_B T}=-\log\left\{\int\Pi_q d\sigma_q \exp(-\Delta
H/k_BT)\right\}=
\log\left\{A(q)^2-B(q)^2\right\}
\label{gibbs}
\end{equation}

The pressure between the two surfaces due to  charge
fluctuations as a function of membrane thickness is given by the
negative derivative of the Gibbs free energy with respect to the thickness:

\begin{equation}
\Pi(d)={k_BT\over A}\sum_q q\exp^{-2qd}{{\delta l\over
<l>}\left(\lambda q+1+{\delta l\over  <l>}\exp^{-2qd}\right)-\left({l_m\over
<l>}\right)^2\over
\left(\lambda q+1+{\delta l\over 
<l>}\exp^{-2qd}\right)^2-\left({l_m\over <l>}\right)^2\exp^{-2qd}},
\label{pressure}
\end{equation}
where we have introduced a Guoy-Chapman length scale: $\lambda={1\over
2\pi <l>\sigma}$. In integral form we find the expression:

\begin{equation}
\Pi(d)=k_BT{1\over 2\pi d^3}\int dx
x^2\exp^{-2x}{{\delta l\over <l>}\left({\lambda\over d}x+1+{\delta l\over
<l>}\exp^{-2x}\right)-\left({l_m\over
<l>}\right)^2\over\left({\lambda\over d}x+1+{\delta l\over
<l>}\exp^{-2x}\right)^2-\left({l_m\over <l>}\right)^2\exp^{-2x}}
\label{pressure integral}
\end{equation}

\subsection {Results and Discussion}
\label {subsec: 2-d res}

The most convenient way to analyze the results of the previous section is by
looking at the various limits of the integral, Eq. \ref{pressure integral}.
We have three dimensionless parameters that determine the behavior of this
integral and thus the $d$ and $\epsilon_w,\epsilon_m$ dependence of the
pressure. The first is the ratio between the two length scales in the
problem:
$${\lambda\over d}= {1\over 2\pi <\!\!l\!\!>\sigma d},$$
which parameterizes the strength of the charging in the membrane relative to
the distance between the surfaces.
The other two parameters are ratios of the dielectric constants and also
their relative difference:
$${\delta l\over <\!\!l\!\!>}={ 2(\epsilon_m-\epsilon_w)\epsilon_m\over
\bar\epsilon^2}\;\;\; {\rm and}\;\;\;
{l_m\over <\!\!l\!\!>}={\epsilon_m\over\bar\epsilon}$$
The first of these two ratios reflects the effect of image charges on
the fluctuation induced interactions, while the second measures the relative
weakening or strengthening of the primary interactions between fluctuations
on the two sides due to the difference in dielectric response of the
material between them.

We have three different parameters with which we find 
two main regimes and one sub-regime. The first regime is reached
when we take the limit ${\lambda\over d}\ll 1$ (high ion density: the average 
distance between ions $\ll\sqrt{d <\!\!l\!\!>}$):
\begin{equation}
\Pi(d)\propto{k_BT\over\pi d^3}\left({\delta l\over
<\!\!l\!\!>}-\left({l_m\over <\!\!l\!\!>}\right)^2\right)\propto-{k_BT\over
d^3}{\epsilon_m\left(2\epsilon_w-\epsilon_m\right)\over\bar\epsilon^2}.
\label{big d uniform}
\end{equation}
The $1/d^3$ behavior remains the same throughout this regime, although the
sign of the pressure changes from being attractive for
$\epsilon_w>\epsilon_m$ (as is expected for a lower dielectric between the
surfaces and is the case for a biomembrane) and even
$\epsilon_m$ slightly bigger than $\epsilon_w$, becoming repulsive only when
the internal dielectric, $\epsilon_m$ is at least twice as big as the
external one, $\epsilon_w$. In this limit the effect of the variation in
the dielectric between the surfaces is just on the size (and eventually the
sign) of the pressure, but the dependence on distance is unaltered
from the uniform case which was described in \cite{sam} as a
fluctuation effect and compared with the Van der Waals attraction also
because of its linear dependence on temperature.

 The next main regime is the opposite
one when
${\lambda\over d}\gg 1$. Here we distinguish between two regimes: The
first is that when the dielectric contrast is not very big (compared with
$\left({l_m\over <l>}\right)^2 {d\over\lambda}$) and in this case the
behavior is, as expected, similar to that found for the uniform case
in this limit\cite{sam}:
\begin{equation}
\Pi(d)\propto-\left({l_m\over <\!\!l\!\!>}\right)^2{k_BT\over
d\lambda^2}\propto-\left({\epsilon_m\over\bar\epsilon^2}\right)^2{\sigma^2\over
d} {e^4\over k_BT}.
\label{small d uniform}
\end{equation}
In this case the pressure is inversely proportional
to the temperature (through the $\lambda$ dependence) and is argued to be a
correlation, rather than a fluctuation, effect
\cite{sam}. The dielectric effects enter in the coefficient $\left({l_m\over
<l>}\right)^2$ and reduce the interaction as the internal dielectric (lipid)
becomes smaller then the external one (water) and the dielectric contrast
increases. However as this contrast increases another effect becomes
important: the effect of the image charges which dominate 
when
$|{\delta l\over <l>}|$ is not small compared with
$\left({l_m\over <l>}\right)^2 {d\over\lambda}$, and we find:
\begin{equation}
\Pi(d)\propto{\delta l\over <\!\!l\!\!>}{k_BT\over
d^2\lambda}\propto-{(\epsilon_w-\epsilon_m)\epsilon_m\over\bar\epsilon^3}
{\sigma
e^2\over d^2}.
\label{small d non uniform}
\end{equation}
Here once again we find that the interaction will change sign when the
internal and external dielectrics reverse roles.
However, the dominant  effect
is that the power law changes from $d^{-1}$ to $d^{-2}$, and therefore for 
smaller $d$ this effect becomes more
important than the previous result, Eq. \ref{small d uniform}. Note that in 
this regime the pressure is
independent of T and is therefore neither pure fluctuation nor correlation
effect. Moreover, it is linearly dependent on the surface charge density,
$\sigma$, (and not quadratically) indicating that the correlations lead to an
average charge distribution which is temperature independent and the result
is an interaction between each charge and its effective image charge which
does not include, to first order, the rest of the mobile charges. 

Because the membrane thickness is typically of order $40\AA$ the limit of
${\lambda\over d}\gg 1$ can only be achieved for very low charging of the
membrane and in this limit it might not be strong enough to compete with the 
Van
der Waals interaction; in any case the stronger power dependence on $d$ might not be
easy to detect. However, note that if we reverse the membrane and water
roles, and we look instead at the inter-membrane interactions say in a stack,
we find that this last result might be more important. Because the
inter-membrane distances in stacks can be relatively small this limit is
easily achieved even in moderately charged membranes (one of every 5-6
lipids is charged). Because the dielectric constants are now reversed, the
ratios change but we remain in this last limit where the reflections
dominate the interaction. Moreover, due to the reversal of the dielectrics,
the interaction (between membranes across the water) is repulsive and
therefore the interplay with the Van der Waals attraction becomes more
interesting. It is especially meaningful in this case because unlike the
lipid material in the membrane, in some experimental set-ups water can flow
in and out of the stack and therefore the stack separation can be more
effectively controlled by this interaction.

In summary, we have shown that fluctuation induced interactions are strongly
dependent on the dielectric properties of the system not only quantitatively
but also qualitatively. The lower dielectric constant of lipid will reduce
the strength of the interaction between the two surfaces of the membrane but
will also change the scaling with the membrane thickness. When looking at
interactions in a stack the reverse happens: the interaction is enhanced by
a factor of
${\epsilon_w\over\bar\epsilon}\simeq 2$ with respect to the uniform case and
we might also be able to see the effects of dielectric reflections when
looking at the inter-membrane interactions. 

\acknowledgments
We are grateful to Sam Safran and Helmut Schiessel for useful
discussions.  R.M was supported by the National Institute
of Health under award No. 1 F32 GM19971. This work was supported by the National Science
Foundation under awards No. 8-442490-21587 and 8-442490-22213, and partially by the 
MRL program of the National Science Foundation under award NO. 
DMR96-32716.

\begin{figure}
\caption{Schematic of model system of membrane ($\epsilon=2$) in water
($\epsilon=80$). The lipid heads are charged both with positive and negative
charges but the membrane is overall neutral. The dashed lines indicate the
virtual surfaces where the image charges show up.  Because there are two
dielectric discontinuities, there are infinitely many such surfaces at equal
distances, $d$, apart.}
\label{autonum}
\end{figure}


\begin{references}
\bibitem{text1}Another difference
enters through the actual $d$ dependence of $B(q)$ through the denominator
that includes the term
$u^2\exp(-2qd)$. However, this exponential term does not have a strong
influence on the resulting inter-surface interaction and therefore on the
equilibrium behavior.
\end{references}
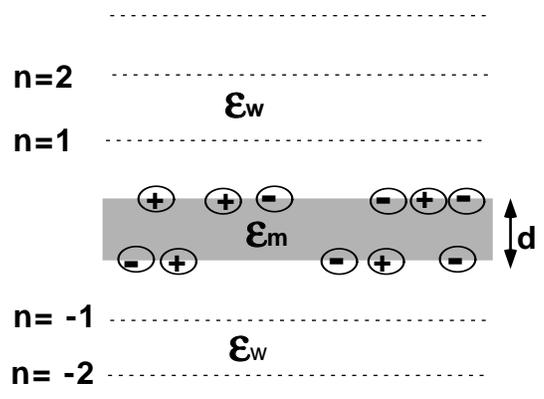
\end{document}